\documentclass[draft]{agujournal2019}
\usepackage{url} 
\usepackage{lineno}
\graphicspath{{.././figures/}}
\journalname{Geophysical Research Letters}

\begin{document}

%
%


\title{The effects of crustal magnetic fields and solar EUV flux on ionopause formation at Mars}

%
%




\authors{F. Chu\affil{1}, Z. Girazian\affil{1}, D. A. Gurnett\affil{1}, D. D. Morgan\affil{1}, J. Halekas\affil{1}, A. J. Kopf\affil{1}, E. M. B. Thiemann\affil{2}, and F. Duru\affil{3}}


\affiliation{1}{Department of Physics and Astronomy, University of Iowa, Iowa City, IA 52242, USA}
\affiliation{2}{Laboratory for Atmospheric and Space Physics, University of Colorado Boulder, Boulder, CO 80303, USA}
\affiliation{3}{Department of Physics, Coe College, Cedar Rapids, IA 52402, USA}




\correspondingauthor{F. Chu}{feng-chu@uiowa.edu}




\begin{keypoints}
\item The Martian ionopause is found to have an average altitude of $363 \pm 65$ km
\item The ionopause altitude has seasonal variations in response to the solar EUV flux
\item The ionopause rarely forms over strong crustal magnetic fields 
\end{keypoints}

%
%


\begin{abstract}
We study the ionopause of Mars using a database of 6,893 ionopause detections obtained over 11 years by the MARSIS (Mars Advanced Radar for Subsurface and Ionosphere Sounding) experiment. The ionopause, in this work, is defined as a steep density gradient that appears in MARSIS remote sounding ionograms as a horizontal line at frequencies below 0.4 MHz. We find that the ionopause is located on average at an altitude of $363 \pm 65$ km. We also find that the ionopause altitude has a weak dependence on solar zenith angle and varies with the solar extreme ultraviolet (EUV) flux on annual and solar cycle time scales. Furthermore, our results show that very few ionopauses are observed when the crustal field strength at 400 km is greater than 40 nT. The strong crustal fields act as mini-magnetospheres that alter the solar wind interaction and prevent the ionopause from forming.
\end{abstract}

\section{Introduction}
\label{sec:intro}

In the absence of a strong global-scale magnetic field, the upper atmosphere of Mars interacts directly with the incident solar wind plasma and interplanetary magnetic field (IMF). The highly conducting ionosphere acts as a barrier to the solar wind flow, leading to a formation of several plasma boundaries. For instance, located downstream of the bow shock is a sharp transition layer called the Magnetic Pileup Boundary (MPB). It separates the magnetosheath from the Magnetic Pileup Region (MPR) where the IMF piles up in front of the planetary obstacle \cite{crider_observations_2002, bertucci_mgs_2004, bertucci_structure_2005}. The photoelectron boundary (PEB), existing between the MPB and the ionosphere, is identified by a drop of $\textup{CO}_2$-photoelectron fluxes in the electron energy spectrum \cite{garnier_martian_2017}. About 200 km below the PEB is the ionopause that marks the transition from the hot, magnetized solar wind plasma to the cold, dense ionospheric plasma \cite{nagy_plasma_2004, han_discrepancy_2014}.


Depending on the definition, the exact location of the ionopause varies in different studies. For example, \citeA{elphic_observations_1980} defined the Venusian ionopause at the location where the ionospheric thermal pressure is balanced by the normal component of the solar wind dynamic pressure. \citeA{han_discrepancy_2014} considered the Martian ionopause to occur where the total electron density drops below a threshold of $10^3$ $\textup{cm}^{-3}$. \citeA{vogt_ionopause-like_2015} identified the ionopause through a sharp decrease in the total ion density by at least a factor of 10 over an altitude range of at most 30 km. In this paper, the ionopause is identified as a steep density gradient that appears in MARSIS (Mars Advanced Radar for Subsurface and Ionosphere Sounding) remote sounding ionograms as a horizontal line at frequencies below 0.4 MHz.




Due to the low ionospheric thermal pressure and presence of the highly-localized crustal magnetic fields at Mars, previous studies suggest that the behavior of the Martian ionopause is rather different from the ionopause at Venus. For example, as the peak thermal pressure in the Mars ionosphere rarely exceeds the solar wind dynamic pressure, the ionopause at Mars is predicted to be a less sharp boundary than Venus \cite{luhmann_characteristics_1987}. Furthermore, the ionopause is observed much less often at Mars and its altitude has a weaker dependence on solar zenith angle (SZA) \cite{duru_steep_2009, han_discrepancy_2014, vogt_ionopause-like_2015}. Over the past few decades, a number of studies have been dedicated to the investigation of the ionopause at Venus \cite{brace_dynamic_1980, elphic_stability_1984, brace_structure_1991, fox_ionosphere_1997, futaana_solar_2017}; however, the investigation of the Martian ionopause still remains limited. 


The MARSIS instrument on board the Mars Express (MEX) spacecraft is a low-frequency radar sounder designed to perform both subsurface and ionospheric soundings \cite{picardi_marsis_2004}. Prior to MARSIS, the ionospheric plasma profiles at Mars were extensively obtained using radio occultation techniques \cite <e.g.,>{zhang_post-pioneer_1990, luhmann_near-mars_1991}. Most recently, the Mars Atmosphere and Volatile EvolutioN (MAVEN) spacecraft has provided accurate in-situ measurements of the electron density with its Langmuir probe \cite{ergun_dayside_2015, jakosky_mars_2015}, but MAVEN can only observe the ionopause twice during each orbit \cite{mendillo_maven_2015, vogt_ionopause-like_2015}. MARSIS remote sounding, on the other hand, is capable of detecting the ionopause continuously from hundreds of kilometers above, providing a unique and powerful tool to study the properties of the ionopause at Mars.

In this paper, we present the first study on the geographic locations of the Martian ionopause in relation to the crustal magnetic fields. The investigation is based on a large sample size of 10,693 orbits from June 2005 to May 2017. We also report the dependence of the ionopause altitude on SZA and the solar extreme ultraviolet (EUV) flux. The paper is organized as follows: section~\ref{sec:observations} gives a description of the ionopause observations through remote radar sounding, section~\ref{sec:results} presents the results, section~\ref{sec:discussion} provides a discussion, and section~\ref{sec:conclusions} gives conclusions of the paper.

\section{Ionopause Observations through Remote Sounding}
\label{sec:observations}


The MARSIS instrument is carried by the MEX spacecraft that has a quasi-polar orbit with inclination of $86.35^\circ$ and 6.75 h period \cite{chicarro_mars_2004}. MARSIS is composed of a 40 m dipole antenna, a 7 m monopole antenna, a radio transmitter, a receiver, and a digital signal processing system \cite{picardi_marsis_2004}. For ionospheric remote sounding, MARSIS sends a short radio pulse from 0.1 to 5.4 MHz and detects any echoes that are reflected from the ionosphere \cite{gurnett_radar_2005}. The reflection occurs because electromagnetic waves cannot propagate in a plasma when their frequency is below the electron plasma frequency given by
\begin{equation}
\label{eq:frequency}
f_\textup{\scriptsize{pe}}=\frac{1}{2\pi}\sqrt{\frac{n_\textup{\scriptsize{e}}e^2}{\varepsilon _0m_\textup{\scriptsize{e}}}},
\end{equation}
where $n_\textup{\scriptsize{e}}$ is the electron density, $e$ is the electron charge ,$m_\textup{e}$ is the electron mass, and $\varepsilon _0$ is the permittivity of free space \cite{gurnett_introduction_2005}. When the transmitter frequency exceeds the maximum plasma frequency, the radio pulse can sometimes penetrate the ionosphere and reflect from the surface of the planet \cite{morgan_solar_2006}. A complete frequency sweep takes 1.257 s and is repeated every 7.54 s. By measuring the time delay $\Delta t$ between the transmission of the pulse and the time that the echo is received, the apparent altitude of the reflection point can be calculated as
\begin{equation}
\label{eq:alt}
h=h_\textup{\scriptsize{MEX}}-\frac{c\Delta t}{2},
\end{equation}
where $h_\textup{\scriptsize{MEX}}$ is the spacecraft altitude and $c$ is the speed of light. It should be noted that apparent altitude is not a ‘‘real’’ altitude scale, in the sense that it has not been corrected for dispersion of the radar pulses that propagate in a plasma \cite{kopf_transient_2008}. Nevertheless, we use the apparent altitude in this study because the dispersion effects are small, due to the low plasma density between the spacecraft and the ionopause. 

\begin{figure*}
\begin{center}
\includegraphics[width=6.69in]{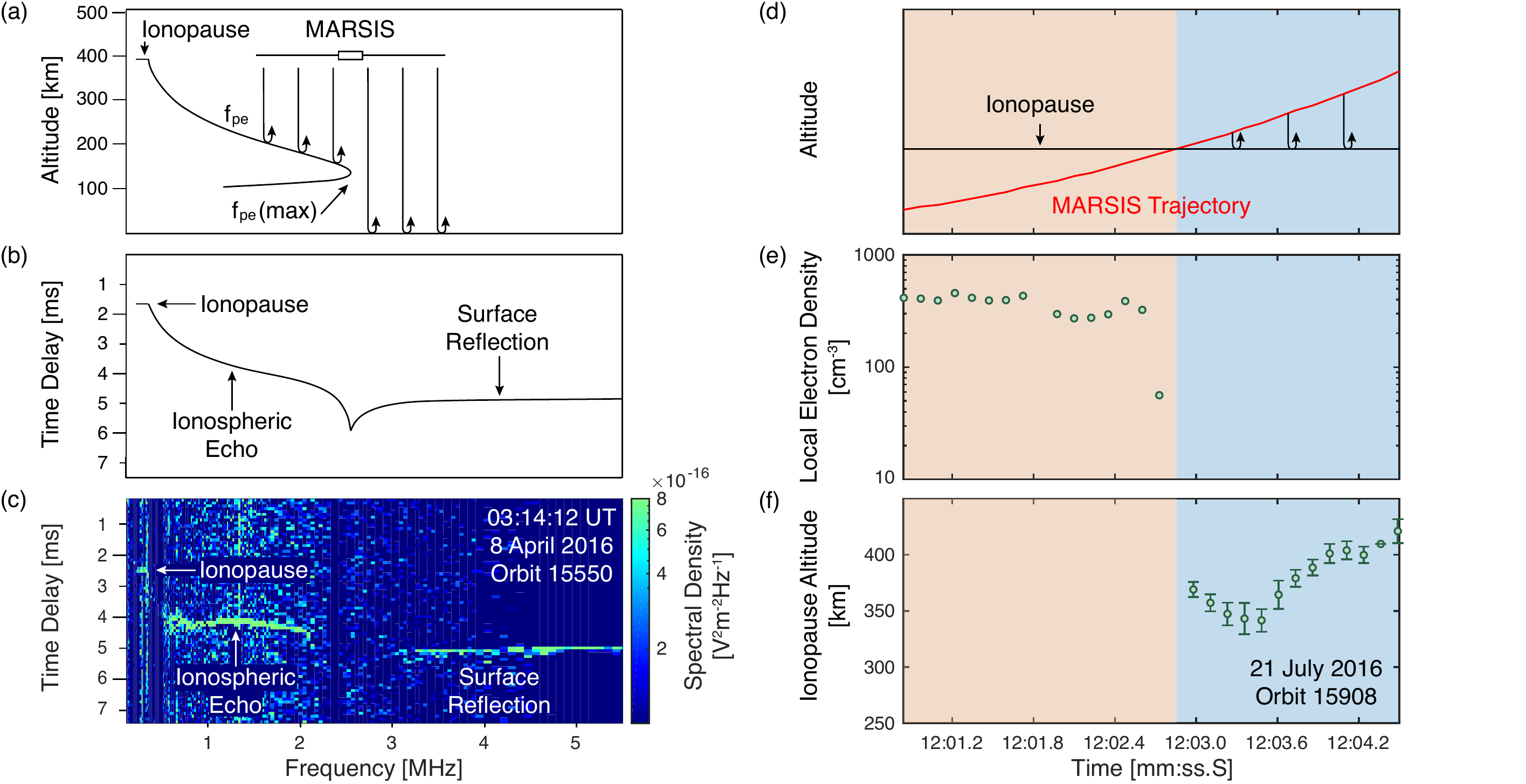}
\caption{Schematic of a typical electron plasma frequency profile as a function of the altitude in the Martian ionosphere is shown in (a) and the resulting ionogram is shown in (b). An example of the color-coded MARSIS ionogram from the orbit 2360 on 14 November 2005 is shown in (c). The ionopause can be seen as a horizontal line at frequencies below 0.4 MHz. An example of MEX flying across the ionopause for an outbound pass on 21 July 2016 is shown in (d)--(f). The local plasma density in (e) is detected using in-situ measurements whereas (f) shows the remote sounding data. The sharp electron density gradient observed using in-situ method in (e) provides strong evidence that the horizontal line in (c) is the ionopause. The time on the horizontal axis has a format of mm:ss.S, where mm stands for minutes, ss seconds, and S fractional seconds. Error bars in (f) represent one-standard-deviation uncertainties. Note the rapid irregular variations in the altitude of the ionopause.}
\label{fig:ionogram}
\end{center}
\end{figure*}

MARSIS remote sounding data are displayed in ionograms, which show the intensity of the returning echoes as a function of the pulse frequency and time delay. The schematic of a typical electron plasma frequency profile as a function of the altitude in the Martian ionosphere along with the resulting ionogram is presented in Figures~\ref{fig:ionogram}a--\ref{fig:ionogram}b. An example of the color-coded MARSIS ionogram from an orbit on 14 November 2005 is shown in Figure~\ref{fig:ionogram}c. The ionopause can be seen as a horizontal line at frequencies below 0.4 MHz, where the time delay of the returning echoes is nearly constant, representing a steep density change over a short vertical distance \cite{duru_steep_2009}. In general, the intensities of the returning echoes between the ionopause and the lower ionospheric trace (e.g., 0.35--0.65 MHz in Figure~\ref{fig:ionogram}c) appear relatively weak in the ionogram. This is because in this region when the time delay changes rapidly with altitude, the energy per unit time of the reflections becomes small due to them being spread over a wider range of time delays.

This sharp electron density gradient can also be detected through in-situ measurements as MEX flies across the ionopause \cite{duru_steep_2009}. When MARSIS transmits the radio pulses, intense electrostatic electron plasma oscillations can be excited at the local electron plasma frequency \cite{gurnett_overview_2008}. These oscillations are then picked up by the receiver, resulting in closely spaced vertical harmonic lines in the low frequency region of the ionogram. The harmonic are caused by nonlinear distortion in the receiver \cite{andrews_determination_2013}. The local plasma density, therefore, can be determined from the frequency of the electron plasma oscillations using equation (\ref{eq:frequency}). An example of MEX flying across the ionopause for an outbound pass on 21 July 2016 is shown in Figures~\ref{fig:ionogram}d--\ref{fig:ionogram}f. Strong local plasma oscillations in the ionosphere are detected at the beginning of the pass (colored in brown) when the spacecraft is below the ionopause. The local electron density slowly decreases and then suddenly a steep density drop is observed as MEX crosses the ionopause. Once the spacecraft is above the ionopause (region colored in blue), the local electron density becomes undetectable and the ionopause echo from remote sounding starts to appear in the ionogram. The example presented in Figures~\ref{fig:ionogram}d--\ref{fig:ionogram}f provides strong evidence that the horizontal line observed in the low frequency region of the ionogram is the ionopause.

A potential issue in identifying the ionopause arises from the ionopause echo sometimes being obscured by the electron plasma frequency harmonics or electron cyclotron echoes in the ionogram \cite{duru_steep_2009}. The later was initially thought to be a bigger problem when remote sounding takes place over the strong crustal magnetic field regions, which could result in undercounting of the actual number of ionopauses. However, after a careful examination, we conclude that MARSIS is fully capable in ionopause detections over the strong crustal field regions, because both the harmonics and electron cyclotron echoes do not exist in the ionogram any more when the spacecraft is above the ionopause, where the local plasma density at the spacecraft location is well below 100 $\textup{cm}^{-3}$. More details and discussions about identifying the ionopause can be found in the supporting information.

\section{Results}
\label{sec:results}

In this study, we only focus on the dayside ionopauses that have SZA less than 100 degrees. We have examined 1,767,529 ionograms over 10,693 orbits from 22 June 2005 to 6 May 2017 to search for ionopause signatures by eye, however, ionopauses are only observed in 6,893 ionograms. The occurrence rate on an orbit by orbit basis is 9\% (958/10,693), which is comparable to the 18\% reported in \citeA{duru_steep_2009}. Our estimation here is likely to be a lower limit of the occurrence rate, as the harmonics caused by the local plasma oscillations sometimes obscure the ionopause echoes in ionograms. It is also not directly comparable to other studies, such as \citeA{vogt_ionopause-like_2015}, which used different ionopause definitions and detection methods.



To understand the mechanisms that control the formation of the Martian ionopause, we have investigated the dependence of the ionopause altitude on SZA and solar EUV radiation. We have also studied the influence of the crustal magnetic fields on the global ionopause occurrence at Mars. Since the apparent altitude in equation~(\ref{eq:alt}) is not corrected for dispersion, the error in the ionopause apparent altitude becomes larger as the distance between MEX and the ionopause increases. For this reason, in sections~\ref{subsec:SZA}--\ref{subsec:season} we only consider the cases where the spacecraft altitude is below 700 km during remote sounding.

\subsection{Dependence of Ionopause Altitude on SZA}
\label{subsec:SZA}

\begin{figure*}
\begin{center}
\includegraphics[width=6.69in]{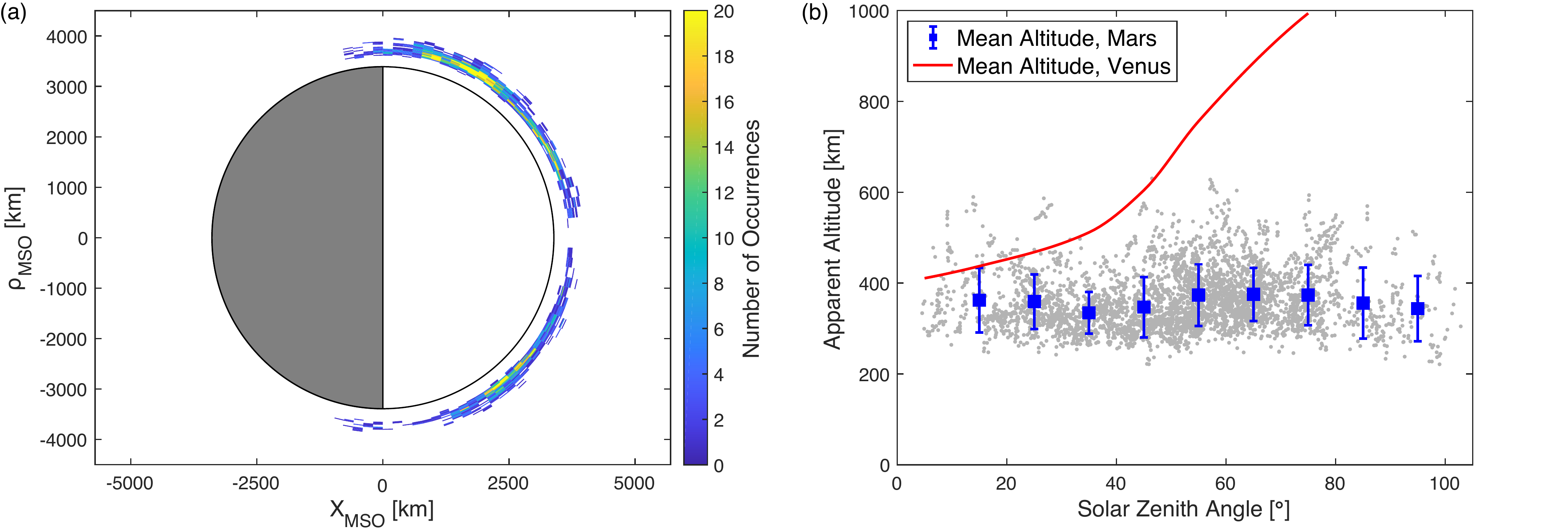}
\caption{Ionopause altitudes from 2005 to 2017 as a function of SZA at Mars are shown in (a) and (b). Figure (a) is plotted in the Mars-Solar-Orbital (MSO) coordinate system, where the $X$-axis points from Mars toward the Sun, $Y$-axis points to the opposite direction of Mars’ orbital velocity, and $Z$-axis completes the right-handed coordinate system. The $\rho$-axis lies in the $Y$-$Z$ plane and is multiplied by the sign of the $Z$-component to differentiate the ionopause between northern and southern hemisphere. The data in (a) are binned every 20 km in the radial direction ($r=\sqrt{\rho^2+X^2}$) and $3^\circ$ in SZA. The mean ionopause altitude averaged over each $10^\circ$ SZA bin is shown in blue squares in (b). Error bars represent one-standard-deviation uncertainties. The red curve marks the mean ionopause altitude at Venus, based on the in-situ measurements from the Pioneer Venus Orbiter's Electron Temperature Probe (OETP). The Venusian ionopause plotted in (b) is defined the same way as the ionopause in this study, a steep electron density gradient on the top of the ionosphere \cite{phillips_asymmetries_1988}.}
\label{fig:alt}
\end{center}
\end{figure*}

At Venus, as SZA increases, the normal component of the solar wind dynamic pressure decreases, causing the ionopause to rise in order to maintain pressure balance \cite{brace_dynamic_1980}. To explore whether similar trends in the ionopause altitude occur at Mars, we show the ionopause apparent altitude as a function of SZA in Figures~\ref{fig:alt}a--\ref{fig:alt}b. We find that 96\% of the ionopauses occur at altitudes between 300 km and 430 km. The average altitude is $363 \pm 65$ km, similar to the previous reported values \cite{han_discrepancy_2014, vogt_ionopause-like_2015} but lower than the 450 km in \citeA{duru_steep_2009}. Furthermore, the occurrence of the ionopause is observed to fall off sharply at 250 km. It appears in Figure~\ref{fig:alt}b that the Martian ionopause altitude does not significantly increase with SZA, which is strikingly different from the ionopause at Venus, as shown in red. This SZA trend in the ionopause altitude at Mars is consistent with the results shown in \citeA{han_discrepancy_2014}.

\subsection{Solar EUV Flux Variations in Ionopause Altitude}
\label{subsec:season}

At Venus, the ionopause forms where the ionospheric thermal pressure balances the magnetic pressure of the MPR. As the solar EUV radiation is the primary energy source responsible for creating and heating the ionosphere \cite{brace_structure_1991}, the altitude of the Venusian ionopause varies over the solar cycle \cite{kliore_solar_1991}. During solar minimum, when the ionospheric thermal pressure is low, the ionosphere is often magnetized and the majority of the ionopause altitudes lie between 200 and 300 km; however, at solar maximum, the ionopause can rise to altitudes between 300 km and more than 1000 km \cite{kliore_solar_1991}. If pressure balance also holds at the Martian ionopause, a similar behavior is expected. Although Mars differs from Venus in that there may also be a seasonal variation owing to the high eccentricity of Mars’ orbit.

\begin{figure*}
\begin{center}
\includegraphics[width=6.69in]{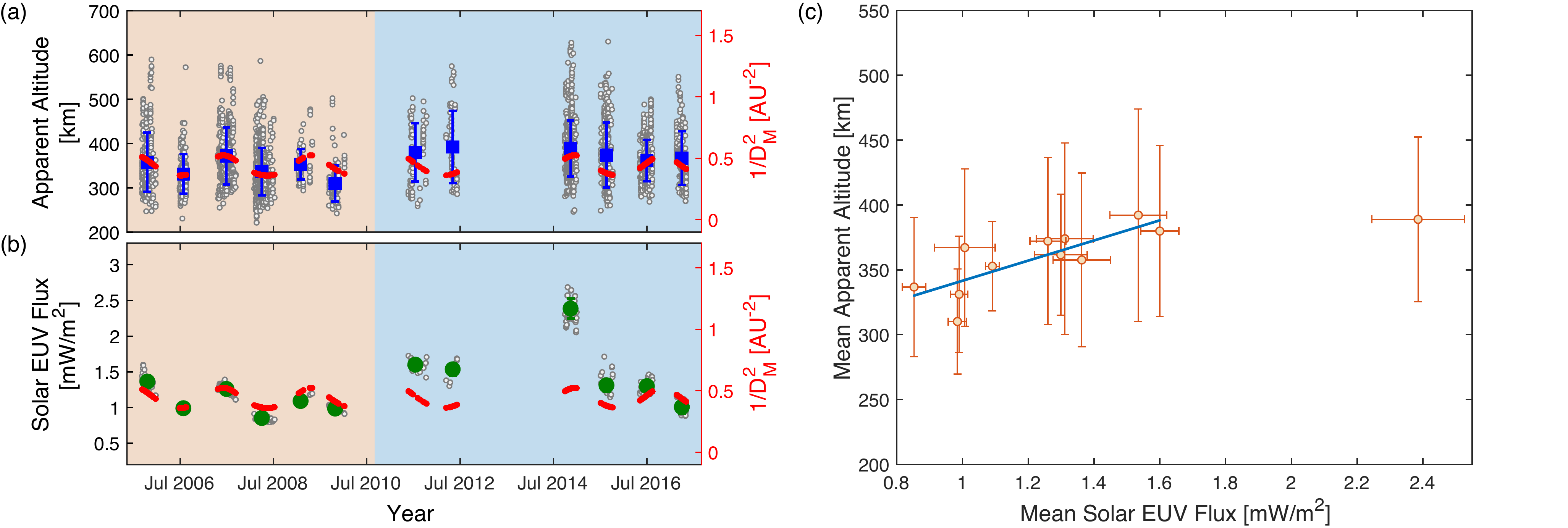}
\caption{(a) Time series of the ionopause altitudes from 2005 to 2017. The right axis shows $1/D_\textup{\scriptsize{M}}^2$, where $D_\textup{\scriptsize{M}}$ is the Mars-Sun distance in AU. The blue square represents the mean ionopause altitude in each cluster of data points, which is naturally separated based on the observation dates. (b) Time series of 12-year solar EUV flux along with $1/D_\textup{\scriptsize{M}}^2$. The mean solar EUV flux is shown in green circles. (c) Correlation between the mean ionopause altitude and mean solar EUV flux. The best fit for the mean solar EUV flux less than 1.6 $\textup{mW/m}^2$ is also plotted here to guide the eye. Error bars in (a)--(c) represent one-standard-deviation uncertainties.}
\label{fig:season}
\end{center}
\end{figure*}

The time series of the ionopause altitudes from 2005 to 2017 is presented in Figure~\ref{fig:season}a. For comparison, the timeline of $1/D_\textup{\scriptsize{M}}^2$ is also shown in the same plot, where $D_\textup{\scriptsize{M}}$ is the Mars-Sun distance in AU. As is most easily noticed in the brown region, the mean ionopause altitude appears to be correlated with $1/D_\textup{\scriptsize{M}}^2$, implying a seasonal variation in the Martian ionopause. The solar EUV flux over the same time period is shown in Figure~\ref{fig:season}b. The data used here are derived from integrating the $\textup{CO}_2$ ionizing component of daily-averaged EUV spectra between 0.5-91.2 nm \cite{girazian_dependence_2013, girazian_empirical_2015}. The EUV spectra are computed using the methods developed for the Flare Irradiance Spectral Model-Mars (FISM-M) \cite{thiemann_maven_2017}. The variation of the EUV flux in Figure~\ref{fig:season}b results from a combination of the changing Mars-Sun distance and the 11-year solar cycle. The blue regions of Figures~\ref{fig:season}a--\ref{fig:season}b may suggest that the ionopause altitude also has a solar cycle trend, being higher during solar maximum than solar minimum.


The relationship between the mean ionopause altitude and the mean solar EUV flux is shown in Figure~\ref{fig:season}c. The correlation coefficient between these two quantities is 0.71, indicating a relatively strong EUV flux effect in ionopause altitude. Depending on the level of the solar EUV flux, the ionopause altitude varies between 340 km and 390 km.

\subsection{Effects of Crustal Magnetic Fields on Ionopause Formation}
\label{subsec:fields}



\begin{figure*}
\begin{center}
\includegraphics[width=6.69in]{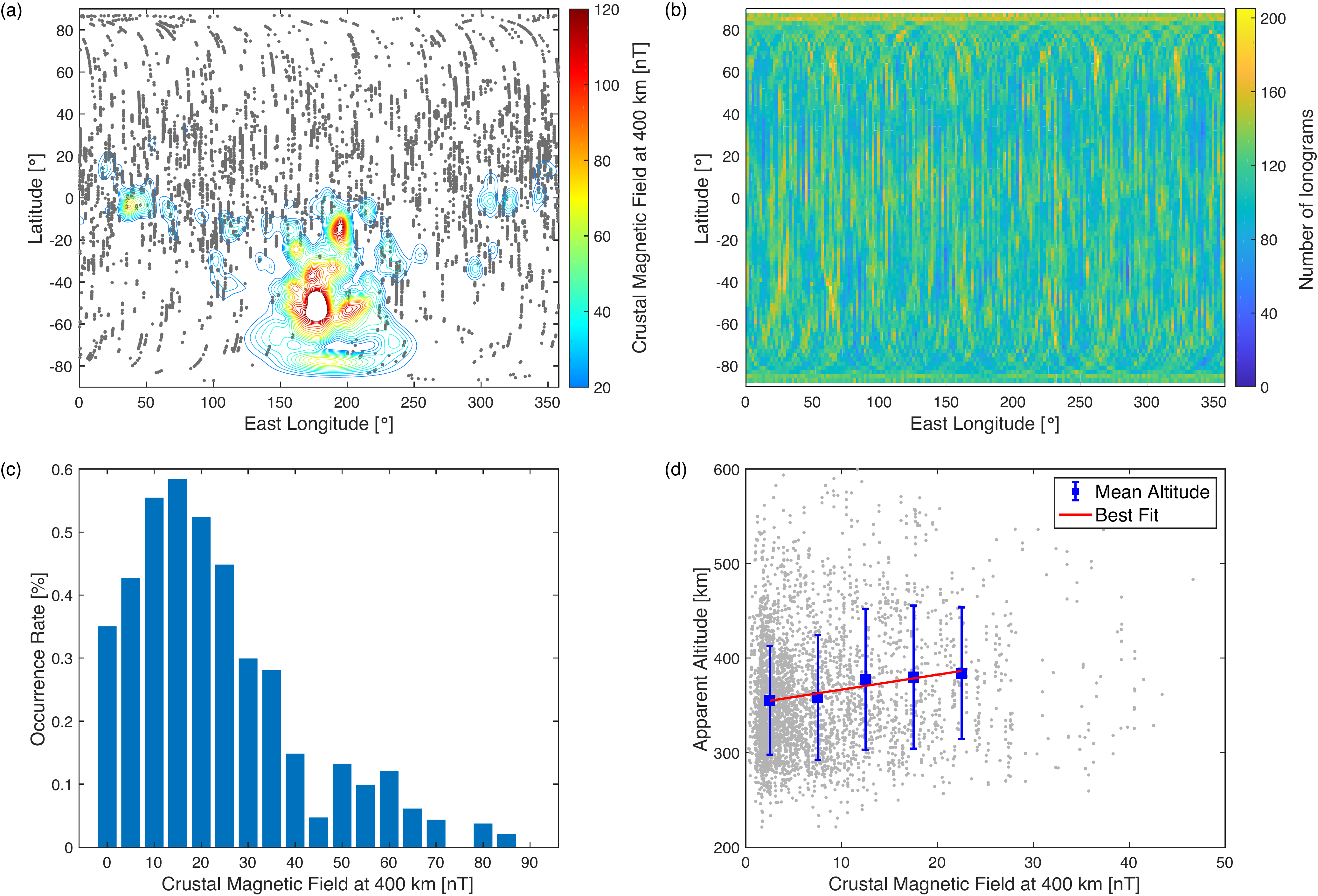}
\caption{(a) Geographic locations of the ionopause in relation to the crustal magnetic field strength at 400 km. (b) Coverage map of MEX in the period of data collection from 2005 to 2017. (c) Histogram of the ionopause occurrence rate as a function of the crustal field strength. (d) Scatter plot of the ionopause apparent altitude as a function of the crustal magnetic field strength at 400 km. The mean ionopause altitudes are shown in blue squares. Error bars represent one-standard-deviation uncertainties.}
\label{fig:B}
\end{center}
\end{figure*}

Unlike Venus, the presence of the crustal magnetic fields at Mars adds complexity to the solar wind interaction and perhaps the formation of the ionopause \cite{nagy_plasma_2004}. In Figure~\ref{fig:B}a we have shown the geographic locations of the ionopause in relation to the crustal magnetic field strength at 400 km \cite{morschhauser_spherical_2014}. The coverage map of MEX during the data collection period is plotted in Figure~\ref{fig:B}b. To remove the observational bias in the data, the ionopause occurrence rate is computed by normalizing Figure~\ref{fig:B}a with Figure~\ref{fig:B}b. The result of this procedure, a histogram of the ionopause occurrence rate as a function of the crustal field strength, is shown in Figure~\ref{fig:B}c. It is clear in both Figure~\ref{fig:B}a and \ref{fig:B}c that the ionopause preferentially occurs above weak crustal fields, and almost never forms near the strongest crustal field region in the southern hemisphere. In fact, most of ionopauses are detected in locations where the crustal field strength at 400 km is less than 40 nT. Figure~\ref{fig:B}a also shows that more ionopauses tend to occur in the northern hemisphere, which can be potentially explained by the strong crustal magnetism in the southern hemisphere.



A scatter plot of the ionopause apparent altitude as a function of the crustal field strength is shown in Figure~\ref{fig:B}d. As the apparent altitude is not corrected for dispersion, for the same reason mentioned earlier, we only consider the cases where the spacecraft altitude is below 700 km during remote sounding. Based on the best fit of the mean ionopause altitudes, it is found that the crustal magnetic fields can increase the altitude of the ionopause by $1.58 \pm 0.34$ km per nT. This result is similar to the one mentioned in \citeA{duru_steep_2009}, where the ionopause altitude is reported to be raised by 25--60 km over strong crustal field region. However, \citeA{vogt_ionopause-like_2015} did not find this trend in their study, possibly because they had a much smaller data set or a different detection method. Similar crustal magnetic field effects are observed as well in other plasma boundaries at Mars \cite{mitchell_probing_2001, crider_influence_2004, edberg_statistical_2008}.

\section{Discussion}
\label{sec:discussion}

From a comparative planetology view, we have addressed three interesting differences between the ionopause properties at Venus and Mars. First, at Venus, the ionopause altitude increases with SZA, which is consistent with the ``flaring'' observed at the bow shock and MPB \cite{futaana_solar_2017}. This behavior is expected when one considers pressure balance across the interface. At Mars, similar trends are found in plasma boundaries such as the bow shock, MPB, ion composition boundary, and pressure balance boundary \cite{vignes_solar_2000, crider_observations_2002, edberg_statistical_2008, matsunaga_statistical_2017, ramstad_solar_2017, gruesbeck_three-dimensional_2018, halekas_structure_2018}. However, as shown in Figure~\ref{fig:alt}b, the flaring of the ionopause is much weaker, suggesting that pressure balance may not be the only mechanism that controls the formation of the ionopause at Mars.

\begin{figure*}
\begin{center}
\includegraphics[width=6.69in]{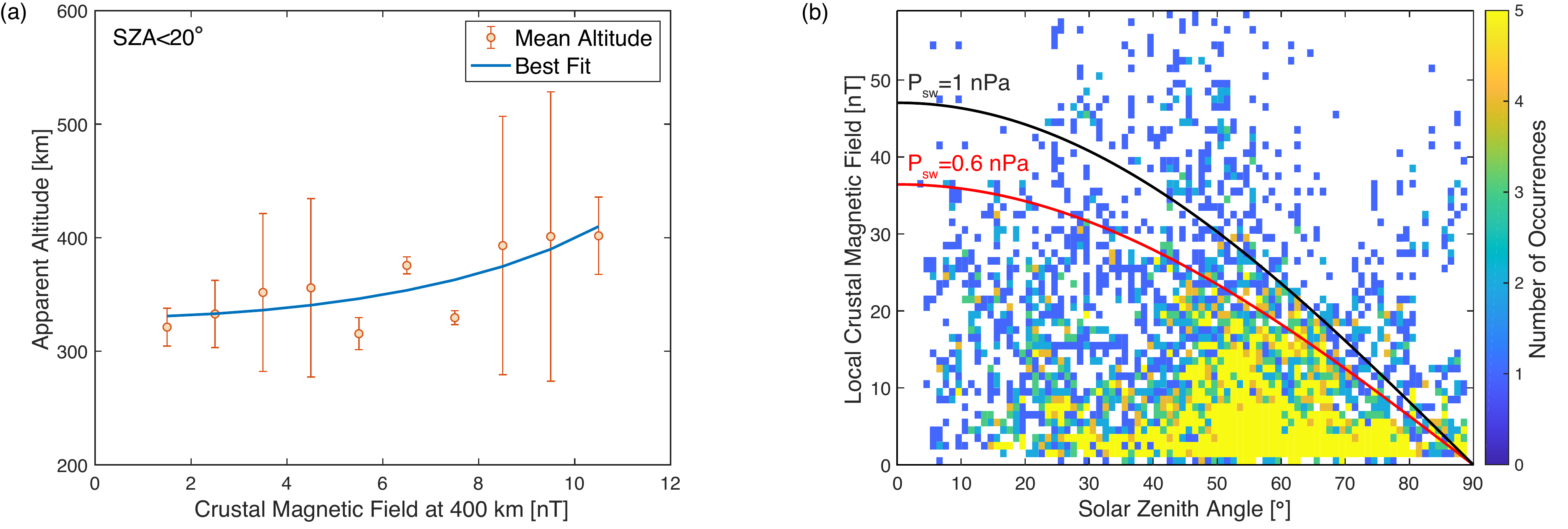}
\caption{(a) Mean ionopause altitude with $\textup{SZA}<20^\circ$ as a function of the crustal magnetic field strength at 400 km. Error bars represent 1$\sigma$ within the averaging bins. The best fit curve is shown in blue. The electron density and temperature used in the fit are $n_\textup{\scriptsize{e0}}=1 \times 10^3$ $\textup{cm}^{-3}$ and $T_\textup{\scriptsize{e0}}=3000$ K at $\sim$ 350 km \cite{ergun_dayside_2015}. (b) Ionopause occurrences as a function of SZA and the tangential component of the crustal field at the location of the ionopause. The red and black curves represent $B_\textup{\scriptsize{max}}$ in equation~(\ref{eq:Bmax}) for typical solar wind dynamic pressures ($P_\textup{\scriptsize{sw}}$). These curves mark the threshold above which crustal fields have enough pressure to hold off the solar wind by themselves. More than 97\% of the detected ionopauses fall below the curves, implying that ionopause formation is inhibited by mini-magnetospheres.}
\label{fig:discussion}
\end{center}
\end{figure*}

Just inside the ionopause, the ionospheric thermal pressure $P_\textup{\scriptsize{th}}$ and the magnetic pressure $P_\textup{\scriptsize{B}}$ should stand off the normal component of the solar wind dynamic pressure $P_{\textup{\scriptsize{sw}}\perp}$ \cite{elphic_observations_1980}
\begin{equation}
\label{eq:balance}
P_{\textup{\scriptsize{sw}}\perp}=P_\textup{\scriptsize{th}}+P_{\textup{\scriptsize{B}}},
\end{equation}
where $P_{\textup{\scriptsize{sw}}\perp}=\alpha P_\textup{\scriptsize{sw}}\cos^2\theta$, $P_\textup{\scriptsize{B}}=B_{\textup{\scriptsize{tot}}}^2/2\mu _0$, $P_\textup{\scriptsize{sw}}$ is the solar wind dynamic pressure, $\theta$ is the angle between the ionosphere surface normal and the flow direction of the upstream solar wind, $B_{\textup{\scriptsize{tot}}}$ is the tangential component of the total magnetic field including the crustal fields and induced field, $\mu _0$ is the permittivity of free space, and $\alpha \approx 0.88$ \cite{crider_proxy_2003}. For the first order approximation where the ionospheric plasma is assumed isothermal, the thermal pressure $P_\textup{\scriptsize{th}}$ is given by \cite{schunk_ionospheres_2009, duru_electron_2019}
\begin{equation}
\label{eq:P}
P_\textup{\scriptsize{th}}=P_0\exp\left ( -\frac{h-h_0}{H} \right ),
\end{equation}
where $P_0=2n_\textup{\scriptsize{e0}}kT_\textup{\scriptsize{e0}}$ (the electrons and ions are assumed to have the same temperature), $k$ is the Boltzmann constant, and $H$ is the scale height. The electron density and temperature at the height of $h_0$ are denoted by $n_\textup{\scriptsize{e0}}$ and $T_\textup{\scriptsize{e0}}$, respectively. If we consider a situation where $P_{\textup{\scriptsize{sw}}\perp}$ is approximately constant, such as over a small SZA range during quiescent solar wind conditions, then any change in $P_\textup{\scriptsize{B}}$ must be accompanied by an equal but opposite change in $P_\textup{\scriptsize{th}}$ in order to maintain pressure balance. If we further ignore the magnetic pressure from the induced field and assume tangential magnetic field at the ionopause, then, given these simplifications, the theoretical ionopause altitude can be written as
\begin{equation}
\label{eq:fit}
h=h_0-H\ln\left ( 1-\frac{B^2}{2\mu _0 P_0} \right ),
\end{equation}
where $B$ is the crustal magnetic field strength and $h_0$ is chosen at $B=0$. Equation~(\ref{eq:fit}) describes how the ionopause altitude should rise to maintain pressure balance in the presence of a crustal field. To test this theory, we select the ionopauses with $\textup{SZA}<20^\circ$ and fit their mean altitudes with the above equation. The result of this procedure is shown in Figure~\ref{fig:discussion}a. The fitted scale height, $H=100 \pm 29$ km, is consistent with other reported values for the topside ionosphere \cite{duru_electron_2008, duru_electron_2019, wu_morphology_2019}. The best fit in Figure~\ref{fig:discussion}a provides evidence that pressure balance plays a role in ionopause formation near the subsolar point and far from strong crustal fields. 

The second interesting difference between the ionopauses at Venus and Mars is the seasonal variability, as shown in Figure~\ref{fig:season}. We interpret this seasonal variation as being driven by the changing EUV flux due to the high eccentricity of Mars' orbit. When the EUV flux increases, the ionospheric thermal pressure at a fixed altitude also increases \cite{sanchezcano_solar_2016}, causing the ionopause altitude to rise in order to satisfy the pressure balance. Such seasonal variations are not observed at Venus because of its near-circular heliocentric orbit. Depending on the level of the solar EUV flux, the ionopause altitude at Mars only varies by about 50 km, whereas the ionopause at Venus has been observed to rise up to more than 1000 km at solar maximum \cite{kliore_solar_1991}.


Last, since Venus is nonmagnetized, the presence of crustal magnetic fields at Mars adds complexity to the formation of its ionopause. We have shown in Figures~\ref{fig:B}a--\ref{fig:B}c that the Martian ionopause rarely forms when the crustal field strength at 400 km is greater than 40 nT. One possible explanation can be seen by examining equation~(\ref{eq:balance}) for the case when $P_\textup{\scriptsize{B}} \gg P_\textup{\scriptsize{th}}$. In this situation, the solar wind is held up solely by the crustal magnetic fields, which act as a mini-magnetosphere where a steep density gradient (ionopause) is no longer required to achieve pressure balance. This interpretation is consistent with the MARSIS observations shown in Figure~\ref{fig:discussion}b, where ionopause occurrences are plotted as a function of SZA and crustal field strength at the ionopause location \cite{morschhauser_spherical_2014}. The red and black curves mark the threshold in the tangential component of the crustal magnetic field $B_\textup{\scriptsize{max}}$, above which the crustal fields can hold off the solar wind by themselves. These thresholds can be determined by setting $P_\textup{\scriptsize{th}}=0$ in equation~(\ref{eq:balance})
\begin{equation}
\label{eq:Bmax}
B_\textup{\scriptsize{max}}=\sqrt{2\mu _0 \alpha P_\textup{\scriptsize{sw}}}\cos\theta.
\end{equation}
The $B_\textup{\scriptsize{max}}$ curves are drawn for typical solar wind dynamic pressures of $P_\textup{\scriptsize{sw}}=0.6$ nPa (50\% percentage) and 1 nPa (75\% percentage) measured by the MAVEN Solar Wind Ion Analyzer (SWIA) \cite{halekas_solar_2015, halekas_structure_2017}. Our calculation shows that 97\% of the ionopauses occur under the 0.6 nPa curve and 98\% under the 1 nPa curve. In other words, almost no ionopauses are observed above $B_\textup{\scriptsize{max}}$ where the crustal magnetic field provides enough pressure to hold off the solar wind, thereby supporting the mini-magnetosphere hypothesis.

\section{Conclusions}
\label{sec:conclusions}

In conclusion, we have studied the Martian ionopause using more than 6,000 MARSIS radar observations from 2005 to 2017. The ionopause is detected in 9\% of orbits at an average altitude of $363 \pm 65$ km. The ionopause altitude is found to have only a weak dependence on SZA, which challenges the expectation that the ionopause forms at the pressure balance point, as is observed at Venus. In addition, we have found that the ionopause altitude varies with the solar EUV flux, and increases near moderately strong crustal magnetic fields. Finally, our analysis shows for the first time that strong crustal magnetic fields impede ionopause formation, adding another way in which crustal magnetism affects the solar wind interaction at Mars.

At Venus, the ionopause is observed to respond dramatically to solar wind pressure variations \cite{brace_dynamic_1980}. Similar behavior is also expected to occur in Martian ionopause. To fully understand the formation of the ionopause at Mars, we will conduct a future investigation to examine how the ionopause of Mars responds to variations in solar wind dynamic pressure using ASPERA-3 (Analyzer of Space Plasma and Energetic Atoms) solar wind measurements \cite{ramstad_solar_2017}.

\acknowledgments
This work was supported by NASA through Contract No. 1560641 with the Jet Propulsion Laboratory. The MARSIS and solar EUV data used in this paper are publicly available through the NASA Planetary Data System (PDS). The Pioneer Venus Orbiter OETP data can be found at \url{https://pds-ppi.igpp.ucla.edu/search/view/?f=yes&id=pds://PPI/PVO-V-OETP-5-IONOPAUSELOCATION-V1.0}. The authors thank Joe Groene for writing the program that obtains MSO coordinates for the MARSIS remote sounding data.


%
%

\bibliography{.././refs}

%
%
%
%
%

\newpage

\end{document}